\documentclass[aip,jcp,reprint,amsmath,amssymb]{revtex4-1}
\usepackage[utf8]{inputenc}
\usepackage[T1]{fontenc}
\usepackage{graphicx}
\usepackage{hyperref}
\usepackage{siunitx}

\renewcommand{\vec}[1]{{\mathrm{\mathbf{ #1}}}}

\graphicspath{{fig/}}

\begin{document}

\title{SPA\textsuperscript{H}M: the Spectrum of Approximated Hamiltonian Matrices representations}

\author{Alberto Fabrizio}
\affiliation{Laboratory for Computational Molecular Design, Institute of Chemical Sciences and Engineering,
\'{E}cole Polytechnique F\'{e}d\'{e}rale de Lausanne, 1015 Lausanne, Switzerland}
\affiliation{National Centre for Computational Design and Discovery of Novel Materials (MARVEL),
\'{E}cole Polytechnique F\'{e}d\'{e}rale de Lausanne, 1015 Lausanne, Switzerland}
\author{Ksenia R. Briling}
\affiliation{Laboratory for Computational Molecular Design, Institute of Chemical Sciences and Engineering,
\'{E}cole Polytechnique F\'{e}d\'{e}rale de Lausanne, 1015 Lausanne, Switzerland}
\author{Clemence Corminboeuf}
\email{clemence.corminboeuf@epfl.ch}
\affiliation{Laboratory for Computational Molecular Design, Institute of Chemical Sciences and Engineering,
\'{E}cole Polytechnique F\'{e}d\'{e}rale de Lausanne, 1015 Lausanne, Switzerland}
\affiliation{National Centre for Computational Design and Discovery of Novel Materials (MARVEL),
\'{E}cole Polytechnique F\'{e}d\'{e}rale de Lausanne, 1015 Lausanne, Switzerland}

\date{\today}

\begin{abstract}
Physics-inspired molecular representations are the cornerstone of similarity-based learning
applied to solve chemical problems. Despite their conceptual and mathematical diversity, this class
of descriptors shares a common underlying philosophy: they all rely on the molecular information
that determines the form of the electronic Schr\"{o}dinger equation. Existing
representations take the most varied forms, from non-linear functions of atom types and positions
to atom densities and potential, up to complex quantum chemical objects directly injected into the ML architecture.
In this work, we present the Spectrum of Approximated Hamiltonian Matrices
(SPA\textsuperscript{H}M) as an alternative pathway to construct quantum machine learning
representations through leveraging the foundation of the electronic Schr\"{o}dinger equation itself:
the electronic Hamiltonian. As the Hamiltonian encodes all quantum chemical information at once,
SPA\textsuperscript{H}M representations not only distinguish
different molecules and conformations, but also different spin, charge, and electronic states. As a
proof of concept, we focus here on efficient SPA\textsuperscript{H}M representations built from the
eigenvalues of a hierarchy of well-established and readily-evaluated ``guess'' Hamiltonians.
These SPA\textsuperscript{H}M representations are particularly compact and efficient for kernel evaluation
and their complexity is independent of the number of different atom types in the database.
\end{abstract}

\maketitle

\section{Introduction}

Modern machine learning (ML) techniques are at the forefront of an unprecedented methodological shift
affecting virtually all fields of chemistry.\cite{HL2021,DB2021,JTBSN2021,UCSGPSTM2021,M2021}
Regardless of the chosen application or
algorithm, the predicting power of artificial intelligence in chemistry is ultimately related to
the choice of a molecular representation, \textit{i.e.} of a numerical descriptor encoding all the
relevant information about the chemical system.\cite{BKC2013,HL2016,MGBOCC2021}

The crucial role of representations is mirrored by the intensive work that has been dedicated to
finding ever more reliable and widely applicable fingerprints.\cite{HL2016,MGBOCC2021} Although there are effectively
infinite ways to input the information about a molecule into a machine learning algorithm,
conceptually molecular representations could be subdivided into well-defined macro categories.
Chemoinformatics descriptors are a comprehensive set of fingerprints that relies either on string-based fingerprints,
such as SMILES\cite{W1988,WWW1989} and SELFIES,\cite{KHNFA2020}
or on readily available and descriptive properties, such as the number of aromatic
carbon atoms, the shape index of a molecule, and its size,\cite{KLK1996,K2000,TC2000,TC2009,DTME2020}
which are usually chosen using an \textit{a priori} knowledge about their correlation with the
specific target.\cite{GVAOLDS2017} A second class of chemical representations
has been introduced very recently, relying on artificial neural networks (and no human input) to
infer suitable descriptors for the learning exercise.\cite{SUG2021}
Finally, physics-based or quantum machine learning (QML) representations include all those fingerprints
inspired by the fundamental laws of physics that govern molecular systems, in particular, the laws
of quantum mechanics and the basic laws of symmetry.

As physics-based representations are rooted in fundamental laws, they are directly applicable to
any learning task, ranging from the regression of molecular properties to revealing the
relationship between molecules in large chemical databases. Although existing quantum machine
learning representations have drastically different mathematical forms and physical motivations,
they all share the same starting point: the position (and often the type) of the atoms in real
space. This choice is not arbitrary and it is intimately related to the connection
between (static) molecular properties and the electronic Hamiltonian $\hat{H}$.

For a fixed nuclear configuration, the information about all the electronic properties of a
molecule is contained in the many-body electronic wavefunction $\Psi(\vec{x}_1,...,\vec{x}_n)$,
as defined by the Schr\"{o}dinger equation. Since the electronic Hamiltonian defines
$\Psi(\vec{x}_1,...,\vec{x}_n)$, the molecular information necessary to fix $\hat{H}$ is in
principle sufficient for a non-linear model to establish a one-to-one relationship with any electronic property.
The expression for
all the universal (\textit{i.e.} non-molecule specific) terms of the Hamiltonian (\textit{e.g.}
kinetic energy) only requires the knowledge of the total number of electrons ($N$). In contrast, the
external potential (the electron-nuclear attraction potential) also depends on the position of the
nuclei $\{\vec{R}_I$\} and their charges $\{Z_I\}$.\cite{SO1989} Under the assumption of charge
neutrality (\textit{i.e.} $N=\sum_I Z_I$), $\vec{R}_I$ and $Z_I$ uniquely fix the form of the
Hamiltonian and thus represent the only required information to characterize the electronic
wavefunction and electronic properties.

Since no two different molecules have the same Hamiltonian, any representation that relies upon
$\vec{R}_I$ and $Z_I$ is guaranteed to satisfy the injectivity requirement of machine learning,
\textit{i.e.} there must be a one-to-one map between the representation of a molecule and its
properties. Nonetheless, injectivity is not the only condition necessary for efficient and
transferable learning. A representation must encode the same symmetries as the target property upon
any transformation of real-space coordinates (equivariance): rotation, reflection, translation, and
permutation of atoms of the same species.\cite{BKC2013,GSD2017,GWCC2018}

To organize in separate groups all the existing physics-based representations,
it is fundamental to define a metric for the classification.
Among all the possibilities, it is useful for the purpose of this work to classify representations
according to the way they use and transform their molecular inputs.

One well-established methodology is to build representations using atom-centered continuous basis functions
from an input containing the type and the position of the nuclei.
This choice is the common denominator of a series of representations
such as the Behler--Parrinello symmetry functions,\cite{BP2007,B2011,ZHWCE2018}
the smooth overlap of atomic positions (SOAP),\cite{BKC2013,GWCC2018}
the overlap fingerprint of Goedecker and coworkers,\cite{ZAFSFRGSGWG2016}
the $N$-body iterative contraction of equivariant features (NICE),\cite{NPC2020}
and the atomic cluster expansion (ACE).\cite{D2019a,D2019b,DBCDEOO2019}

Other descriptors such as the many-body tensor representation (MBTR),\cite{HR2017}
permutation invariant polynomials (PIPs),\cite{BMBJB2004, BB2009,BBCCCFHKSSWX2010,XB2010,JG2013}
and graph-based representations\cite{PA2011} rely on the transformation of the structural input
into a system of internal coordinates and use directly this information to establish similarity measures.

A third possibility is to build representations as fingerprints of potentials.
This family includes the Coulomb Matrix (CM),\cite{RTML2012,RRL2015} the Bag of Bonds (BoB),\cite{HBRPLMT2015}
(atomic) Spectrum of London and Axilrod--Teller--Muto potential [(a)SLATM],\cite{HL2020}
the long-distance equivariant (LODE) representation,\cite{GC2019} FCHL18,\cite{FCHL2018} and FCHL19.\cite{CBFL2020}

More recently, sophisticated neural network architectures, such as OrbNet,\cite{QWAMM2020,CSQOSDBAWMM2021}
have shown that it is possible to use even more complex quantum chemical objects as input features,
such as the tensor representation of quantum mechanical operators
and their expectation values obtained from a converged semi-empirical computation.

In this work, we propose a different approach to designing physically motivated and efficient QML representations.
The Spectrum of Approximated Hamiltonian Matrices (SPA\textsuperscript{H}M)
looks back at the common origin of physics-based representations and uses the electronic Hamiltonian
as the central ingredient to generate an input for machine learning algorithms.
In contrast to standard geometry-based descriptors,
the Hamiltonian encodes all the relevant quantum chemical information at once
and it is able not only to distinguish different molecules and conformations,
but also different spin, charge, and electronic states. Importantly, SPA\textsuperscript{H}M representations
do not require any self-consistent field (SCF) computation, as they all leverage the simplest,
yet powerful, quantum chemical trick: the use of well-established, low-cost ``guess'' Hamiltonians,
which are traditionally used to jump-start the SCF procedure.
These matrices are cheaper to compute than a single SCF iteration
(see Section~\ref{sec:efficiency} for more details on efficiency) and form a controlled hierarchy
of increasing complexity and accuracy that is readily computed for any given molecular system.
As a proof of concept, we focus in this work on
SPA\textsuperscript{H}M representations built from the eigenvalues of the ``guess'' Hamiltonians.
This choice is physically and chemically motivated, naturally invariant under the basic symmetries
of physics, and, in contrast to existing QML representations, include seamlessly the information
about the number of electrons and the spin state of a molecule.
In addition, the choice of eigenvalues results in a small-size representation,
which is particularly efficient for kernel construction and rather independent
of the degree of chemical complexity in the databases.
Eigenvalue-based SPA\textsuperscript{H}Ms are global representations,
which have the benefit to be rather accurate
for molecular properties, but are not as transferable as local representations
and are not applicable to regress local (atomic) targets (\textit{e.g.} atomic partial charges).\cite{BKC2013}
Nonetheless, the SPA\textsuperscript{H}M representations are not restricted to eigenvalues and
could be constructed from other (atom-centered) properties, such as the ``guess'' Hamiltonian
matrix elements, the eigenvectors, and their corresponding density matrices.

\section{Computational methods}

\begin{figure*}[htb]
\centering
\includegraphics[height=0.62\linewidth]{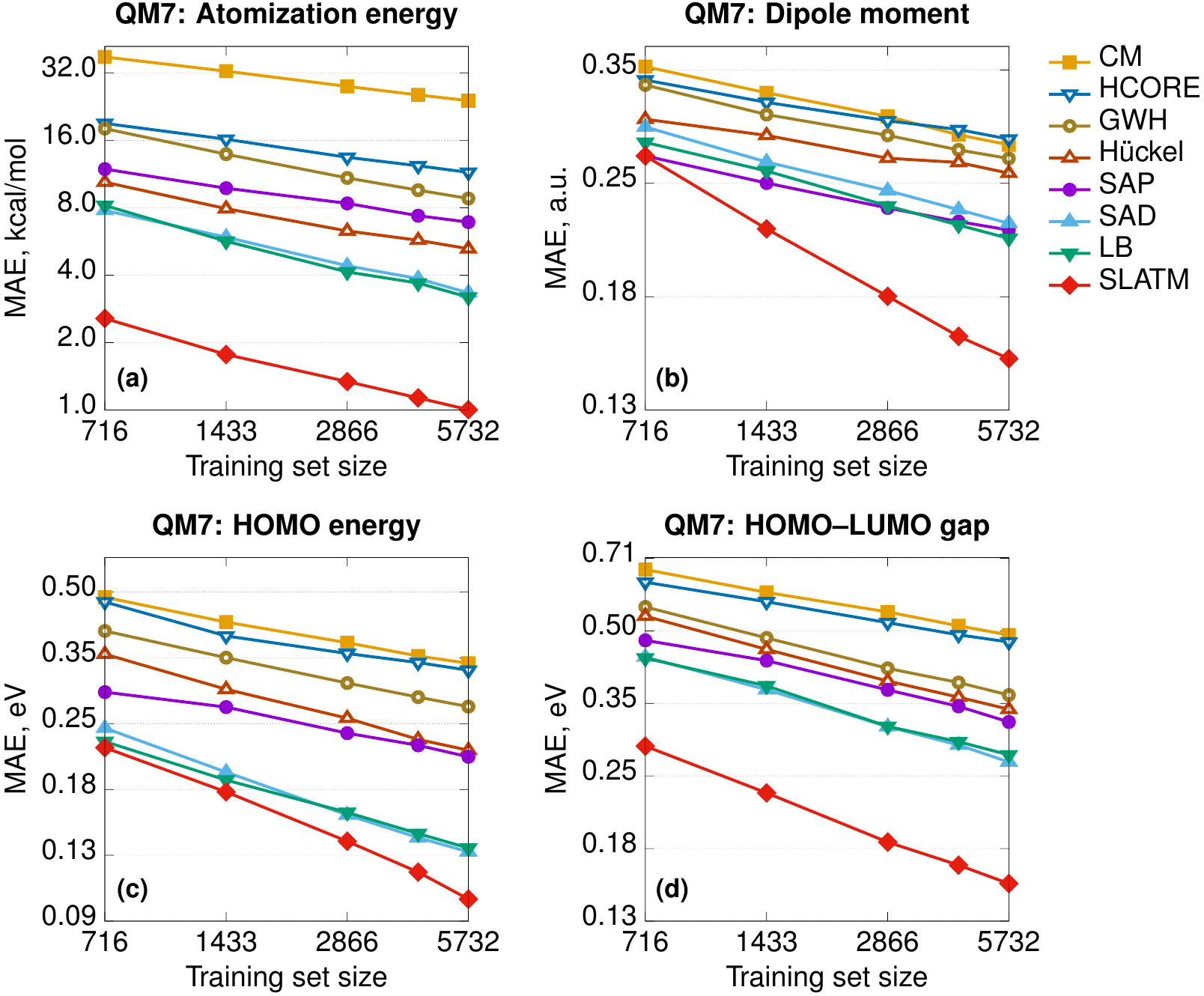}\hfill
\includegraphics[height=0.62\linewidth]{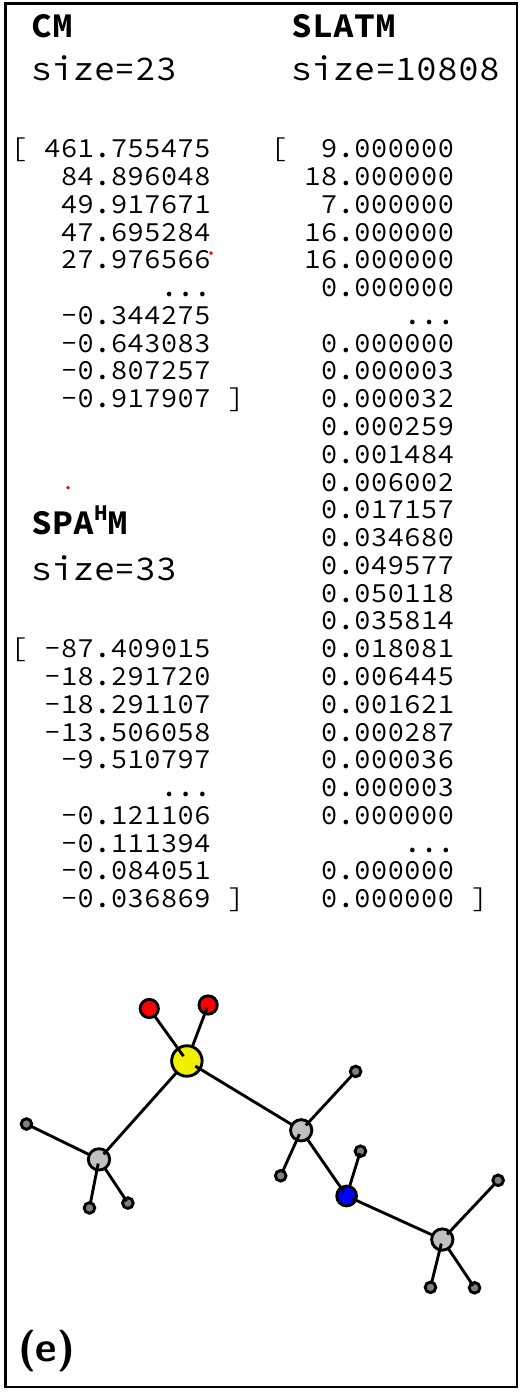}
\caption{\emph{Left.} Learning curves in logarithmic scale of (a) atomization
energies, (b)  dipole moments, (c) HOMO energies, and (d) HOMO--LUMO gaps. The color
code reflects the different representations. \emph{Right.} Illustrative example
of the sizes of the CM, SPA\textsuperscript{H}M, and SLATM representations.
All the Hamiltonians were evaluated in the MINAO\cite{K2013} minimal basis.}
\label{fig:lc}
\end{figure*}

Each molecular set (described in the corresponding subsections of the Results and discussion)
was randomly divided into the training and test sets (80--20\% splits). To minimize the bias
arising from the uneven distribution of molecular size and composition, the learning curves for each
representation were averaged over 5 repetitions of sampling and prediction (error bars are
additionally reported in the Supplementary Information).
While the atomization energies were taken as computed in the original QM7 reference
(PBE0 in a converged numerical basis),\cite{RTML2012} the
other three properties (norm of dipole moment, HOMO energies and HOMO--LUMO gap)
were computed at PBE0\cite{AB1999}/cc-pVQZ\cite{D1989,WD1993} level.
The structure and properties of the molecules in the L11 database were taken as computed in the original reference.\cite{L2011}
The hyperparameters for each representation were optimized with a grid search
using a $5$-fold cross-validation procedure
and the learning curves were computed using random sub-sampling ($5$~times per point).
The optimization and regression code was written in \texttt{Python}
using the \texttt{numpy}\cite{HMWGVCWTBSKPHKBHFWPGSRWAGO2020} and \texttt{scikit-learn}\cite{PVGMTGBPWDVPCBPD2011} libraries.
The \texttt{QML}\cite{CFHBTMV2017} package was used to construct the CM and SLATM representations.
The Gaussian kernel was used for the SLATM representation and the Laplacian kernel for CM and all the SPA\textsuperscript{H}Ms.

The initial guesses were obtained in a minimal basis (MINAO\cite{K2013}).
All quantum chemical computations were made with a locally modified version of \texttt{PySCF}.\cite{S2015,SBBBGLLMSSWC2017}.
The codes used in this paper are
provided in a Github repository at \url{https://github.com/lcmd-epfl/SPAHM}
and are included in a more comprehensive package called
\texttt{Q-stack} (\url{https://github.com/lcmd-epfl/Q-stack}). Q-stack is a library and a collection of stand-alone codes,
mainly programmed in \texttt{Python}, that provides custom quantum chemistry operations to promote quantum machine learning.
The data and the model that support the findings of this study are freely available
in Materials Cloud at \url{https://archive.materialscloud.org/record/2021.221} (DOI:10.24435/materialscloud:js-pz).

The CPU timings were recorded on
24-core CPU servers (2x Intel Xeon CPU E5-2650 v4 @ 2.20GHz),
using one thread.
The code was run with the packages from anaconda-5.2.0 (python-3.6)
together with numpy-1.16.4, pyscf-2.0.0a, and qml-0.4.0.
The user time was measured with the \texttt{getrusage()} system call and averaged over eight runs.

\section{Results and discussion}

\subsection{Learning curves}
\label{sec:lc}

To assess the ability of the SPA\textsuperscript{H}M representations to learn and their overall
accuracy, we trained a kernel ridge regression (KRR) model on the QM7~database\cite{BR2009,RTML2012}
to target four quantum chemical properties.  Each of these quantities has been chosen as it
is representative of a particular category. Atomization energies are
both routinely used to assess the quality of ML models and represent a broader class of extensive
(\textit{i.e.} size-dependent) thermodynamic properties.\cite{RTML2012,RVBR2012} Dipole moments are traditionally
used in quantum chemistry as proxies of the quality of the wavefunction. The HOMO energies
are intensive (\textit{i.e.} size-independent) quantities. Finally, the HOMO--LUMO gap allows
probing the quality of both frontier orbitals and simultaneously tests the additivity of errors in
the KRR models.

The QM7~dataset\cite{RTML2012} was randomly divided into a training set of 5732 molecules and a
test set containing the remaining 1433 compounds, corresponding to an 80--20\% split.
For each molecule, we constructed the SPA\textsuperscript{H}M representations by diagonalizing
the different ``guess'' Hamiltonians in a minimal basis set (MINAO,\cite{K2013})
and using the sorted occupied eigenvalues as the KRR fingerprint.
The occupied orbital energies carry information about both the atom types (core-orbital
eigenvalues), the general electronic structure of the molecule (core and valence), and the total number of electrons.
In addition, the eigenvalues of a Hamiltonian are naturally invariant under all real-space transformations
(permutation, translation, rotation) and the size of the occupied set is independent of the
choice of the atomic orbital basis.

The learning curves for all the SPA\textsuperscript{H}M representations are reported in Figure~\ref{fig:lc}.
In addition, we report the curves of the original version (eigenvalue) Coulomb matrix (CM)\cite{RTML2012} and SLATM,\cite{HL2020}
as the first has a similar size and diagonalization philosophy as SPA\textsuperscript{H}M and the
second is an example of a widely-used global representation.

In Figure~\ref{fig:lc}, the SPA\textsuperscript{H}M representations are indicated by the type of approximate
Hamiltonian used for their construction. Although it is rather complex to establish a definitive
hierarchy of self-consistent-field guesses, it is always possible to provide a more qualitative
trend based on the amount of physics that each guess includes.
The diagonalization of the core ($\vec{{H}}_{\mathrm{core}}$)
and the generalized Wolfsberg--Helmholz (GWH)\cite{WH1952} Hamiltonian matrices
are the simplest approximations, as they do not try to model any
two-electron term. Building on the GWH guess, the parameter-free extended H\"{u}ckel method uses approximate
ionization potentials as diagonal terms and it is generally more robust.\cite{H1963,L2019} The superposition
of atomic densities (SAD)\cite{AFK1982,AC2001,LZDG2006} is another popular choice that however only produces a density matrix
(DM). Nonetheless, it is rather straightforward to construct a complete Hamiltonian matrix
(including the one- and two-electron terms) by contracting the SAD density matrix with a potential
of choice (Hartree--Fock or any exchange-correlation density functional). We report the SAD
learning curve in Figure~\ref{fig:lc} using the PBE0 potential, as all the properties were computed with
this functional. Finally, the superposition of atomic potentials (SAP)\cite{L2019,LVE2020} and the
Laikov--Briling (LB)\cite{LB2020} guesses use effective one-electron potentials to construct sophisticated,
yet computationally lightweight, guess Hamiltonians.

Besides the internal hierarchy, the accuracy of all the SPA\textsuperscript{H}Ms
is always comprised between SLATM and the eigenvalues of the Coulomb Matrix. While
SLATM consistently outperforms the SPA\textsuperscript{H}M representations, the difference with the
most robust guesses (LB and SAD) is usually much smaller than the accuracy of
the functional itself ($\sim 5$ kcal/mol).\cite{LT2003} Importantly, SLATM is also
three orders of magnitude larger than SPA\textsuperscript{H}M on QM7. The
significant difference in the extent of the representation is crucial from an
efficiency perspective, as the number of features dictates the computational
effort of constructing the kernel matrix for an equal size of the training set.
While lightweight, efficient, and naturally accounting for the charge state of molecules,
we only tested a few well-known Hamiltonians for building SPA\textsuperscript{H}M.
As the performance of SPA\textsuperscript{H}M is
largely independent of the choice of the basis set or potential (see
Supplementary Information), it is necessary to consider alternative strategies to
improve its accuracy. The heavy dependence of the learning on the quality of
the parent Hamiltonian suggests that the construction of ``better guesses'' is
the correct direction. Nonetheless, as discussed in Section~\ref{sec:noise},
\textit{better guesses} does not necessarily mean ``improved quantum chemical
approximate Hamiltonians'' (\textit{i.e.} closer to the converged Fock matrix),
but rather the construction of simpler, systematic Hamiltonians specifically
optimized for the learning task.

Besides the rather simple organic molecules of QM7, we tested the accuracy of the best performing
SPA\textsuperscript{H}M representation (LB) on larger molecules, transition metal complexes, and
conformers. Even in these more challenging chemical situations, SPA\textsuperscript{H}M shows the
same relative performance with respect to existing representation as in QM7 (see Supplementary
Material, Section II).

\subsection{Physics and noise}
\label{sec:noise}

\begin{figure}
\centering
\includegraphics[width=\linewidth]{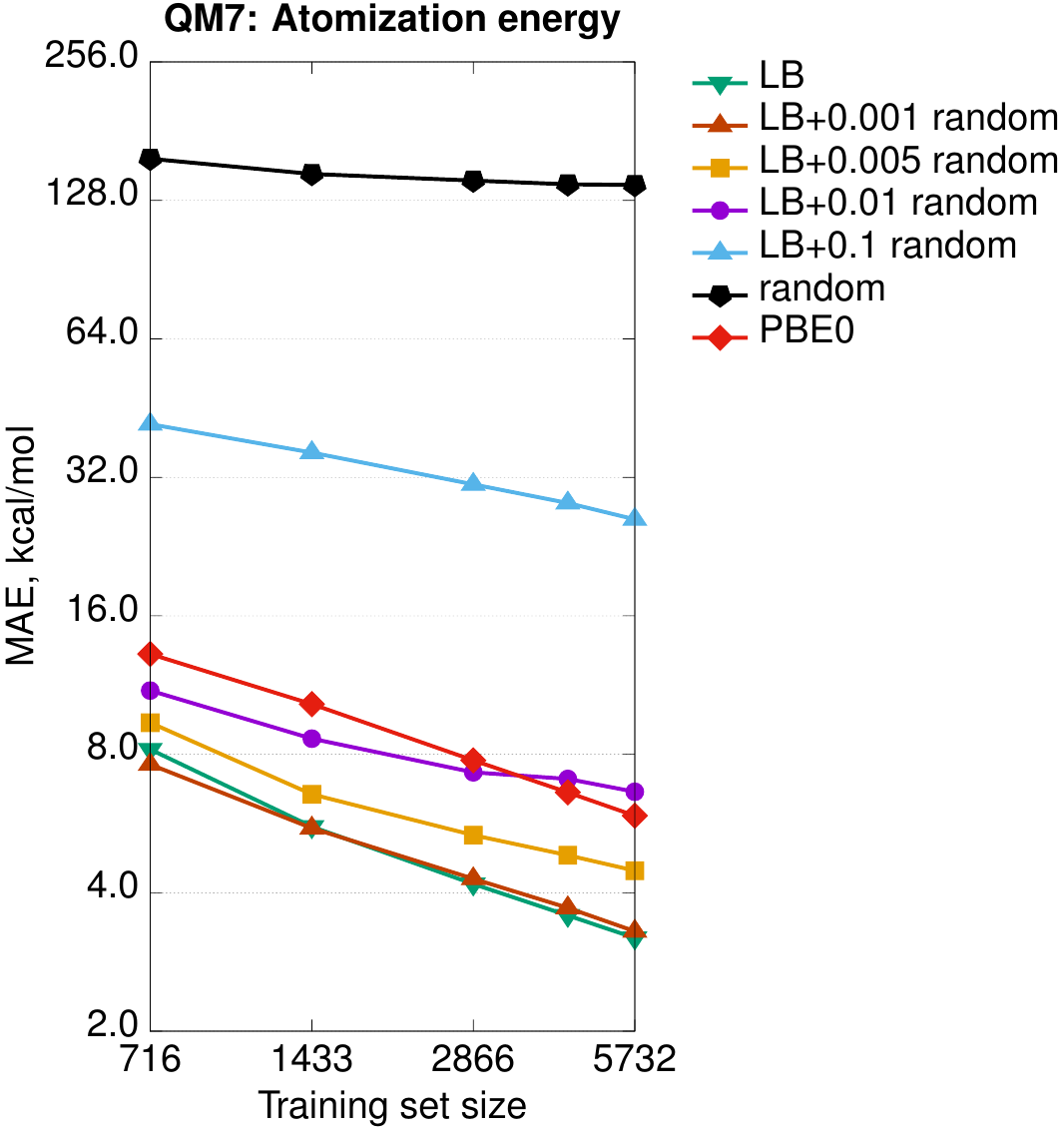}
\caption{Learning curves (in logarithmic scale) of atomization energies for
SPA\textsuperscript{H}M based on the converged PBE0 Fock matrix and on the LB Hamiltonian, with an increasing
(pseudo-)random perturbation added to the representation vector.
The perturbation max. magnitude is reflected in the legend.
For comparison, a learning curve for a fully-random representation is shown.
All the Hamiltonians (including converged PBE0) were evaluated in the MINAO\cite{K2013} minimal basis.}
\label{fig:random}
\end{figure}

\begin{figure*}[htb]
\centering
\includegraphics[width=0.75\linewidth]{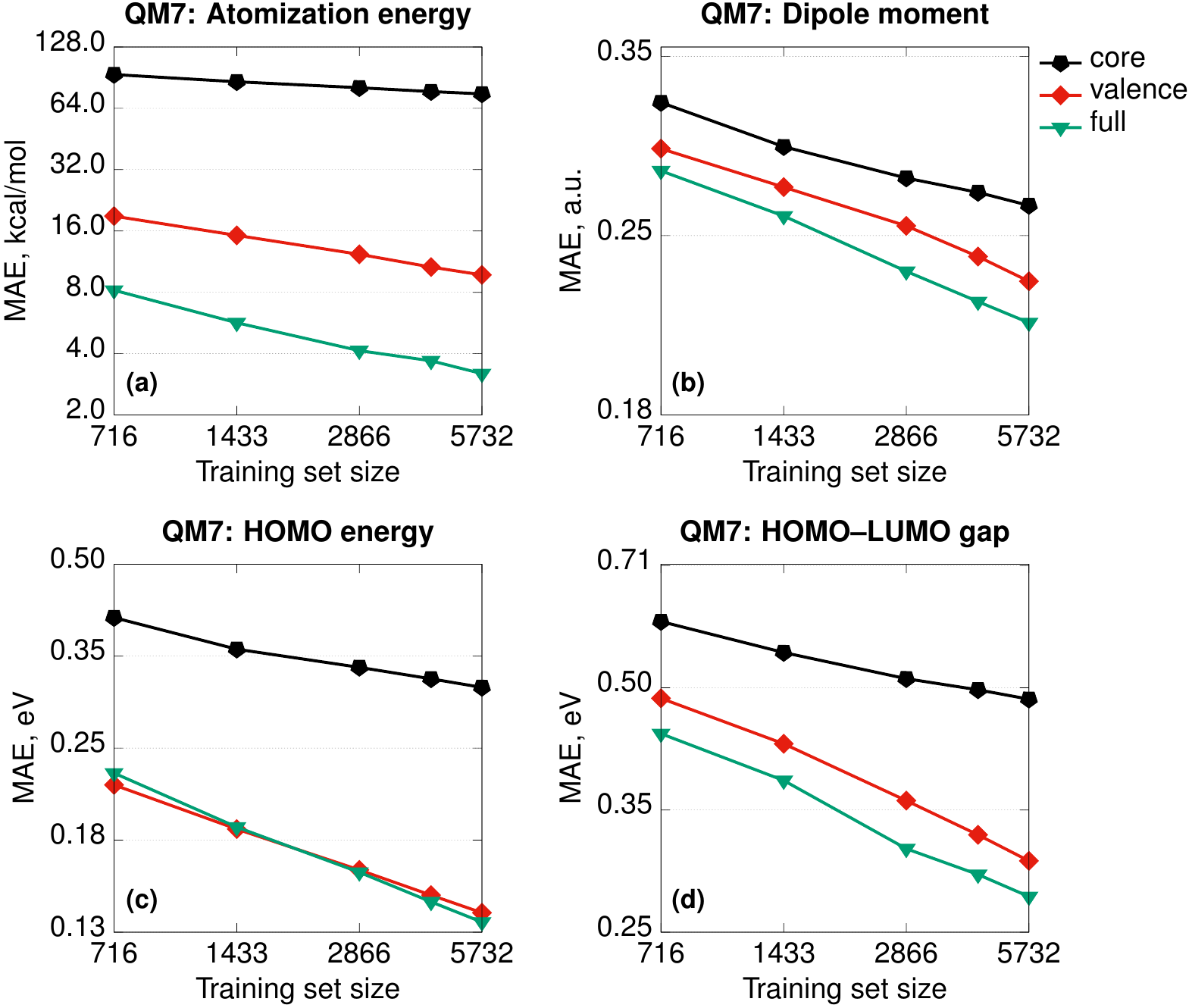}\hfill
\caption{Learning curves (in logarithmic scale) for the SPA\textsuperscript{H}Ms
based on the LB Hamiltonian with core, valence,
and the full occupied eigenvalues sets used as representations.
All the Hamiltonians were evaluated in the MINAO\cite{K2013} minimal basis.
}
\label{fig:corevalence}
\end{figure*}

In general, the accuracy of the different SPA\textsuperscript{H}M representations (Figure~\ref{fig:lc})
follows the same trend as the complexity of the underlying SCF guess. This result seems to suggest
that the more physics is included in the approximate Hamiltonian, the easier is the learning
exercise for the corresponding SPA\textsuperscript{H}M representation. To test the robustness of
this conclusion, we constructed a \textit{test} representation using the converged PBE0 Hamiltonian matrix
(Figure~\ref{fig:random}, label~PBE0). As already mentioned, any representation
based on the converged Fock matrix is both too expensive and worthless for
practical machine learning, since it is always possible to (upon diagonalization) use the converged
wavefunction to compute any desired quantum chemical
property. Nonetheless, this test is essential, as it pushes the physics of
SPA\textsuperscript{H}M to the limit. Figure~\ref{fig:random} shows that PBE0
is not the best representation when regressing the atomization
energies and even some SPA\textsuperscript{H}Ms outperform its accuracy. As the SCF changes the eigenvalues of each molecule
independently from the others, the relationship between the feature vectors also varies
unconcertedly. This sparsification of the data in the representation space effectively decreases
the correlation between the features and the target properties and worsens the learning.

The performance of the converged Fock matrix \textit{versus} more approximated
(and computationally cheaper) Hamiltonians shows that ``more physics'' is not
necessarily the key to better learning. Yet, the relative ordering of the guess
Hamiltonians suggests that higher-quality potentials correlate with the best
representations. The question associated with the relevance of the physics
could be generalized and one could ask if there is the need for any physics at
all or random featurization could lead to the same (or better) results.\cite{CK2018}
In addition, SPA\textsuperscript{H}M representations are so small (33~features for QM7)
with respect to the size of the dataset (7165~molecules),
that the learning could be the result of random correlations between the
features and the target properties. Overall, it is still unclear if any random
(\textit{i.e.} not physically motivated) perturbation of
SPA\textsuperscript{H}M could lead to better learning. To analyze the
behavior of our most robust representation upon random perturbation, we modify
SPA\textsuperscript{H}M-LB by adding an increasing (pseudo-)random
perturbation sampled from a uniform distribution and testing its accuracy on the
QM7 database.

Figure \ref{fig:random} shows that, for the smallest perturbation tested
(magnitude max.~0.001), the original and the modified learning curves are
almost indistinguishable, except for a non-significant difference due to the shuffling of the training set.
As the allowed random noise increases, we observe the systematic and progressive worsening of the learning exercise
towards the limit of physically meaningless random numbers. As the magnitude of
the perturbation increases, the physics in the representation fades, the error
increases, and the learning curves become flatter.
This demonstrates that the performance of SPA\textsuperscript{H}M
is not just a consequence of a random correlation between the feature vectors and the properties.

\subsection{Core and valence}

\begin{figure*}[ht]
\centering
\includegraphics[width=1.0\linewidth]{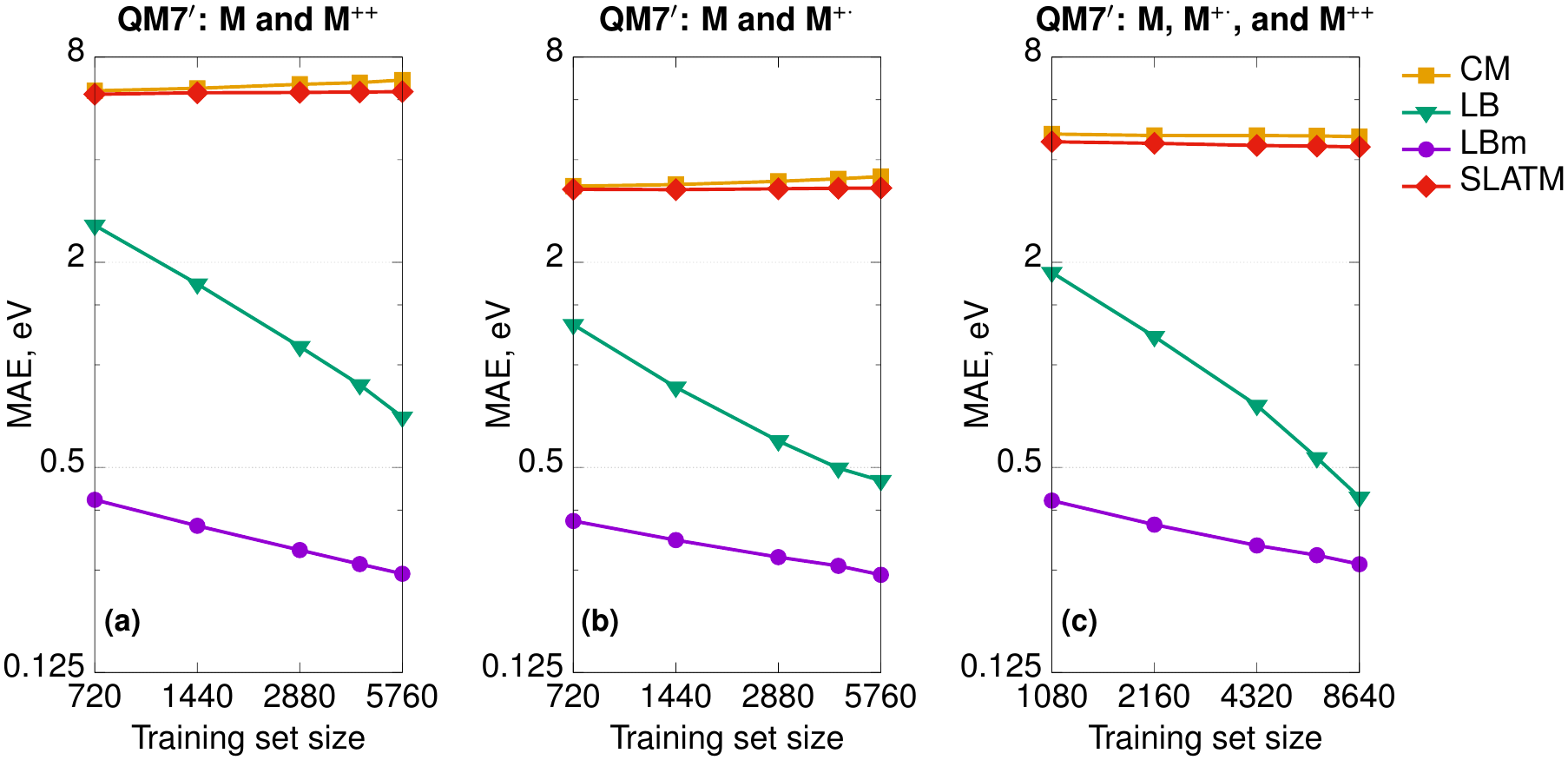}\hfill
\caption{Learning curves (in logarithmic scale) of HOMO energies
for the SPA\textsuperscript{H}Ms based on the LB Hamiltonian (LB and LBm)
compared to CM and SLATM
for the artificial sets including neutral molecules ($\rm M$) taken from QM7
and their double cations ($\rm M^{++}$) and/or radical cations ($\rm M^{+\cdot}$).
}
\label{fig:spincharge}
\end{figure*}

\begin{figure}[ht]
\centering
\includegraphics[width=0.8\linewidth]{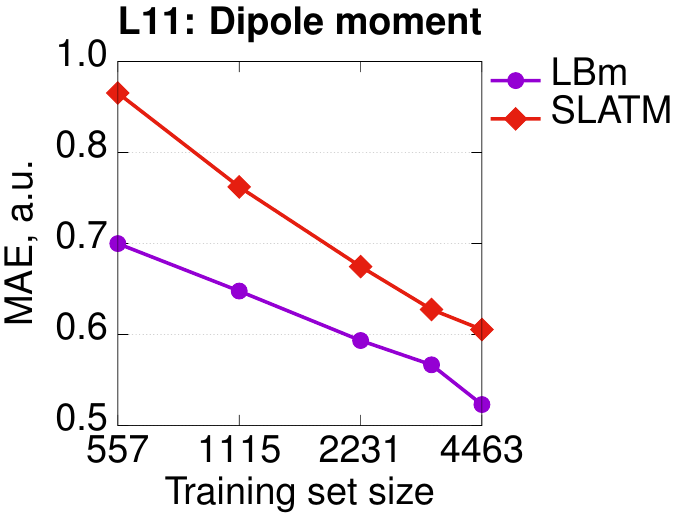}\hfill
\caption{Learning curves (in logarithmic scale) of dipole moments
for the SPA\textsuperscript{H}M based on the LB Hamiltonian (LBm)
compared to SLATM
for the set of Ref.~\citenum{L2011} (L11).
}
\label{fig:npmmes}
\end{figure}

Besides any consideration about the relative importance of physics, the
SPA\textsuperscript{H}M representations are still the eigenvalues of
(approximate) electronic Hamiltonians. As such, it is relevant to try to
rationalize how the different parts of SPA\textsuperscript{H}M contribute to
the learning. As for any set of eigenvalues, it is always possible to divide any
SPA\textsuperscript{H}M representation into its core and valence parts. The core orbital
energies do not vary significantly for the same atom in different molecules and
therefore naturally track the number and the types of nuclei across the
database. In contrast, the valence set is a fingerprint of the chemistry and
the bonding patterns proper to each molecule.

Figure~\ref{fig:corevalence} shows the learning curves for the
SPA\textsuperscript{H}Ms based on the LB Hamiltonian with core, valence,
and the full occupied eigenvalues sets taken as representation vectors.
While the valence set results in consistently better
learning than the core, both are necessary to achieve the overall accuracy of
SPA\textsuperscript{H}M-LB with the exception of the HOMO energies. For the HOMO
eigenvalues, the valence set can be considered an alternative type of
$\mathrm{\Delta}$-learning,\cite{RDRL2015} where the approximate
baseline (approximate HOMO energies) are the input of the kernel itself, rather
than corrected \textit{a posteriori}. Importantly, core orbital energies are
not sufficient information to accurately regress the atomization energies, but
the valence set error is also twice as large as the total
SPA\textsuperscript{H}M. Therefore, while the information about the chemical
bonding is more relevant for the general performance, the information about the
number and the type of nuclei in the molecules is essential to improve the
learning.

\subsection{Spin and charge}

Existing QML representations such as SLATM are computed as non-linear functions
of atom types and internal coordinates.
However, from a quantum chemical perspective, this information is not sufficient to fix the
(ground-state) wavefunction, which also requires knowledge of the number of electrons ($N$).
Omitting the charge (and spin) information is not particularly problematic under the assumption of
electroneutrality in the dataset, \textit{i.e.} the number of electrons in each molecule is exactly
equal to the sum of the nuclear charges ($N=\sum_I Z_I$). Nonetheless, existing geometry-based
representations are not suitable for datasets of molecular systems with different charges (spin states),
since the injectivity rule is violated (the same geometry would correspond to multiple target properties)
or, in the milder case of relaxed geometries, the representation-to-property mapping is not smooth.
By construction and by choice of the electronic Hamiltonians as the key ingredient, SPA\textsuperscript{H}M representations
include naturally both the structural (geometry and atom type) and the electronic (spin and charge)
information and they are applicable with no modification to any molecular database.

To demonstrate the difference in performance between geometry- and Hamiltonian-based
representations on more complex databases, we randomly selected one half of the QM7~set
(3600~molecules) and computed at fixed geometries the properties of the double cations ($\rm M^{++}$)
and radical cations ($\rm M^{+\cdot}$). In this way, we constructed three additional sets:
\emph{(a)} neutral molecules and double cations (7200~molecules, 5760~in the training and 1440~in the test set);
\emph{(b)}~neutral molecules and radical cations (7200~molecules, 5760~in the training and 1440~in the test set); and
\emph{(c)} neutral molecules, double cations, and radical cations (10800~molecules, 8640~in the training and 2160~in the test set).
We set the learning task to predict the HOMO (SOMO for radicals) orbital energies
and report the learning curves in Figure~\ref{fig:spincharge}.
As expected, CM and SLATM fail the learning exercise and the curves are flat for all three sets.

SPA\textsuperscript{H}M representations include the charge information on two separate levels.
First, as we only include the occupied space, the length of the representation changes if we remove
electrons. For instance, the SPA\textsuperscript{H}M representations of neutral molecules $\rm M$
and their double cations $\rm M^{++}$ differ by one entry in length. Second, some of the
approximate potentials at the origin of SPA\textsuperscript{H}M, \textit{e.g.} LB, SAP, and SAD,
are sensitive to electronic information and result in different Hamiltonian matrices when the
number of electrons differs. For instance, the LB Hamiltonian relies on a constraint that fixes the
charge corresponding to the effective Coulomb potential (LBm in Figure~\ref{fig:spincharge}).
While both LB (with no modified potential) and LBm learn, it is evident from
Figure~\ref{fig:spincharge}, that SPA\textsuperscript{H}M-LBm provides more robust predictions,
resulting in errors one order of magnitude smaller than SPA\textsuperscript{H}M-LB at small training set sizes.

The spin-state information is also readily included in SPA\textsuperscript{H}Ms by the same method
used in traditional quantum chemistry: separating the $\alpha$ and $\beta$-spaces and concatenating
$\alpha$ and $\beta$ orbital energies in a matrix of size $N_\alpha\times2$ (the $\beta$-orbitals
column is padded with zeros). Using the Laplacian kernel function, this choice ensures that for closed-shell molecules
the similarity measures are the same as described above when the fingerprint is a single vector of length $N/2$.

\smallskip

Hamiltonian-based representations such as SPA\textsuperscript{H}M outperform geometry-based
representations in every database where electronic information is fundamental to distinguish
molecules. The three sets proposed above are an example, but they are also quite peculiar since we
do not allow geometries to relax.
As a more realistic example of a database where electronic information is essential,
we compare the overall performance of SPA\textsuperscript{H}M (LBm Hamiltonian) and SLATM
on the L11 set.\cite{L2011}
L11 consists of 5579~small molecular systems (single atoms and atomic ions were excluded from the
original set) characterized by a substantial diversity in terms of chemistry, charge, and spin states.
From L11, 4463~molecules were randomly selected as the training set and the remaining 1116 --- as the test set.
To assess the accuracy of the representation, we set the norm of the dipole moment
(for charged systems the origin is chosen to be the geometric center)
as the learning target and report the learning curves in Figure~\ref{fig:npmmes}.
Since there are no identical geometries in the set,
the injectivity rule is not violated for SLATM, and the representation learns.
However, even with relaxed structures, geometry-based representations struggle with a database
containing mixed electronic information since they are not smooth with respect to the target property
(\emph{similar} geometries can correspond to significantly different values),
and SPA\textsuperscript{H}M, incorporating seamlessly the electronic state information, performs better.

\subsection{Efficiency}
\label{sec:efficiency}

The efficiency and computational complexity of SPA\textsuperscript{H}M representations
depend on the choice of the underlying guess Hamiltonian, the simplest guesses (core, GWH) being the fastest to evaluate.
Formally, as the framework requires the diagonalization of a matrix,
the overall complexity in big-O notation is $O(N^3)$ where $N$ is a measure of the system size.
SLATM, despite being a much larger representation (see Fig.~\ref{fig:lc}e),
also scales as $O(N^3)$, since it includes three-body interactions.
Formal complexities for a single molecule are a useful analysis tool,
but they are not always sufficient to characterize the efficiency of the representations on a full dataset.
In this case, practical examples are a more compelling demonstration of the relative merits
of the SPA\textsuperscript{H}M philosophy.

\begin{figure}[ht]
\centering
\includegraphics[width=1.0\linewidth]{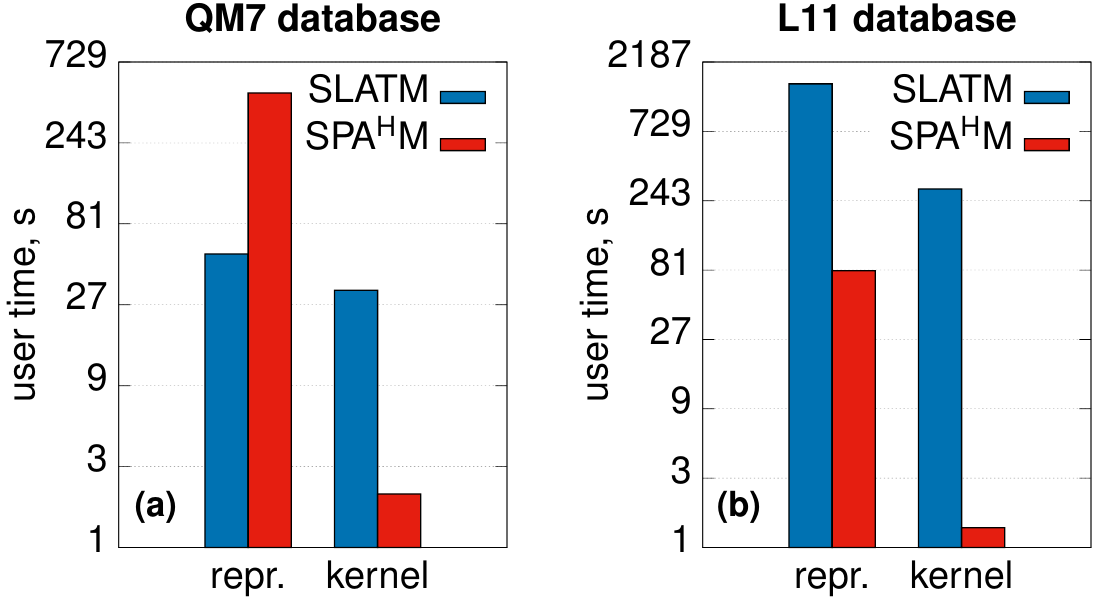}\hfill
\caption{
User times (in logarithmic scale) measured for computing the SLATM and SPA\textsuperscript{H}M representations
and the molecular kernels on \emph{(a)} the QM7 dataset\cite{RTML2012} and
\emph{(b)} the database of Ref.~\citenum{L2011} (L11).
}
\label{fig:timing}
\end{figure}

For this reason, we report in Figure~\ref{fig:timing} the CPU timing for SLATM and SPA\textsuperscript{H}M-LB(m)
on the QM7 and the L11\cite{L2011} databases (more details in the Computational Methods).
For both representations, we recorded the time for building the representation itself
and the time for constructing the kernels.

As SPA\textsuperscript{H}M representations based on occupied orbital eigenvalues are particularly compact,
SPA\textsuperscript{H}M-LB is significantly more efficient than SLATM in the kernel construction for both databases.
This result follows directly from the fact that, for a fixed dataset size,
the complexity of the kernel construction depends only on the number of features
(SLATM: \num{10808} features on QM7 and \num{249255} on L11; SPA\textsuperscript{H}M: 33 on QM7 and 64 on L11).

The relative efficiency of the construction of the representation itself is more sensitive to the dataset.
To better understand the behavior of both SLATM and SPA\textsuperscript{H}M-LB(m)
it is crucial to clarify the composition of each database.
QM7 includes molecules up to 7 non-hydrogen atoms with a limited chemical complexity (H, C, N, O, S).
In contrast, L11 contains all the elements Hydrogen through Bromine (noble gases excluded)
and as such includes a significantly diverse chemistry.
SLATM is faster to evaluate than SPA\textsuperscript{H}M-LB in the QM7 dataset,
as the amount of different many-body types is very limited.
However, on more complex databases with diverse chemistry and atom types,
representations like SLATM become heavy in both computer time and memory
(SLATM binary file occupies 11 GiB of memory \textit{vs} 5.5 MiB of SPA\textsuperscript{H}M).
As the cost of the guess Hamiltonian does not depend on the number of different atom types,
SPA\textsuperscript{H}M representations are extremely efficient for chemically complex molecular databases.

\section{Conclusions}

In this work, we proposed a lightweight and efficient quantum machine learning
representation, capable of naturally accounting for the charge state of molecules
by leveraging the information contained in standard quantum chemical
``guess'' Hamiltonians. Using the QM7 and the L11 databases, we tested the
performance of a hierarchy of representations for a set of four representative
quantum chemical properties. The performance of the SPA\textsuperscript{H}M
representations follows the same qualitative trend as the one describing the
amount of physics encoded in the parent approximate Hamiltonian. Nonetheless,
we also find that the trend stops when pushing the physics to the limit and
using the fully converged Fock matrices to construct the representation. Since
increasingly adding physics is not the roadmap for the potential improvement
of the SPA\textsuperscript{H}M representations, alternative strategies have to be analyzed.
Sharing the same conceptual origin of this work, an alternative strategy for future improvement
consists in representing molecules for ML using the approximate Hamiltonian
matrix itself, its eigenvectors, or the resulting density matrices. Together
with the SPA\textsuperscript{H}Ms, these descriptors form a more comprehensive class
of Hamiltonian-centered fingerprints leveraging the simplest, computationally efficient,
and robust, quantum chemical trick: SCF guesses.

\section*{Author Contributions}
A.F. and K.R.B. performed the computations and developed the software.
A.F. and C.C. designed the representations and conceptualized the project.
All the authors contributed to the writing, reviewing and editing of the manuscript.
C.C. is credited for funding acquisition.

\section*{Conflicts of interest}
There are no conflicts to declare.

\section*{Acknowledgements}
The authors acknowledge Ruben Laplaza and Puck van Gerwen
for helpful discussion and critical propositions.
The authors also acknowledge the National Centre of Competence in Research (NCCR)
``Materials' Revolution: Computational Design and Discovery of Novel Materials (MARVEL)''
of the Swiss National Science Foundation (SNSF, grant number 182892) and
the European Research Council (ERC, grant agreement no 817977).

\bibliography{refs}

\end{document}

\title{{\sc Supplementary Information}\texorpdfstring{\\}{}
\SPAHM: the Spectrum of Approximated Hamiltonian Matrices representations}

\author{Alberto Fabrizio}
\affiliation{Laboratory for Computational Molecular Design, Institute of Chemical Sciences and Engineering,
\'{E}cole Polytechnique F\'{e}d\'{e}rale de Lausanne, 1015 Lausanne, Switzerland}
\affiliation{National Centre for Computational Design and Discovery of Novel Materials (MARVEL),
\'{E}cole Polytechnique F\'{e}d\'{e}rale de Lausanne, 1015 Lausanne, Switzerland}
\author{Ksenia R. Briling}
\affiliation{Laboratory for Computational Molecular Design, Institute of Chemical Sciences and Engineering,
\'{E}cole Polytechnique F\'{e}d\'{e}rale de Lausanne, 1015 Lausanne, Switzerland}
\author{Clemence Corminboeuf}
\email{clemence.corminboeuf@epfl.ch}
\affiliation{Laboratory for Computational Molecular Design, Institute of Chemical Sciences and Engineering,
\'{E}cole Polytechnique F\'{e}d\'{e}rale de Lausanne, 1015 Lausanne, Switzerland}
\affiliation{National Centre for Computational Design and Discovery of Novel Materials (MARVEL),
\'{E}cole Polytechnique F\'{e}d\'{e}rale de Lausanne, 1015 Lausanne, Switzerland}

\date{\today}

\maketitle
\onecolumngrid
\tableofcontents

%%%%%%%%%%%%%%%%%%%%%%%%%%%%%%%%%%%%%%%%%%%%%%%%%%%%%%%%%%%%%%%%%%%%%%%%%%%%%%%
\newpage
\section{Influence of the basis set and the potential on \SPAHM}
%%%%%%%%%%%%%%%%%%%%%%%%%%%%%%%%%%%%%%%%%%%%%%%%%%%%%%%%%%%%%%%%%%%%%%%%%%%%%%%

\begin{figure*}[h]
\centering
\includegraphics[width=0.75 \textwidth]{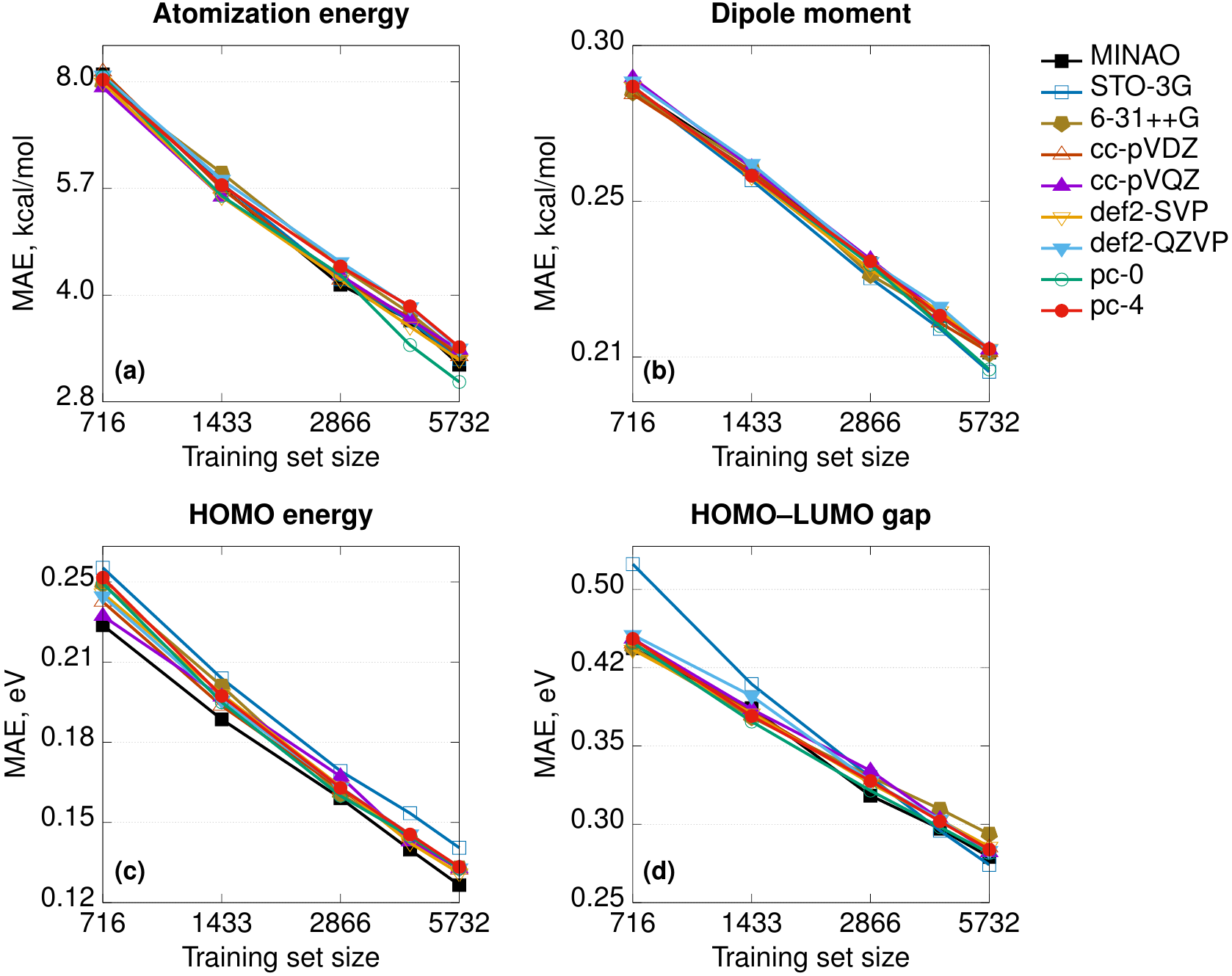}
\caption{
Learning curves for \SPAHM based on the LB Hamiltonian using different basis sets.
}
\label{fig:basis}
\end{figure*}

The size of the electronic Hamiltonian in the atomic orbitals basis depends on
the size of the chosen basis set. Nonetheless, the number of occupied molecular
orbitals, which fixes \SPAHM, only depends on the total number of electrons. In
the main text, we have shown the learning curves of the different \SPAHM
representations using a minimal basis set (MINAO).\cite{K2013}
Figure~\ref{fig:basis} demonstrates that the overall accuracy of \SPAHM is
largely independent of the choice of the atomic basis. The basis sets reported
in the Figure are dramatically different in size and serve distinct conceptual
purposes (\eg converging correlation, capturing polarization effects,
describing diffuse electron clouds \textit{etc.}). At full training, no
significant difference is observable across the entire basis set spectrum.

\begin{figure*}[p]
\centering
\includegraphics[width=0.75 \textwidth]{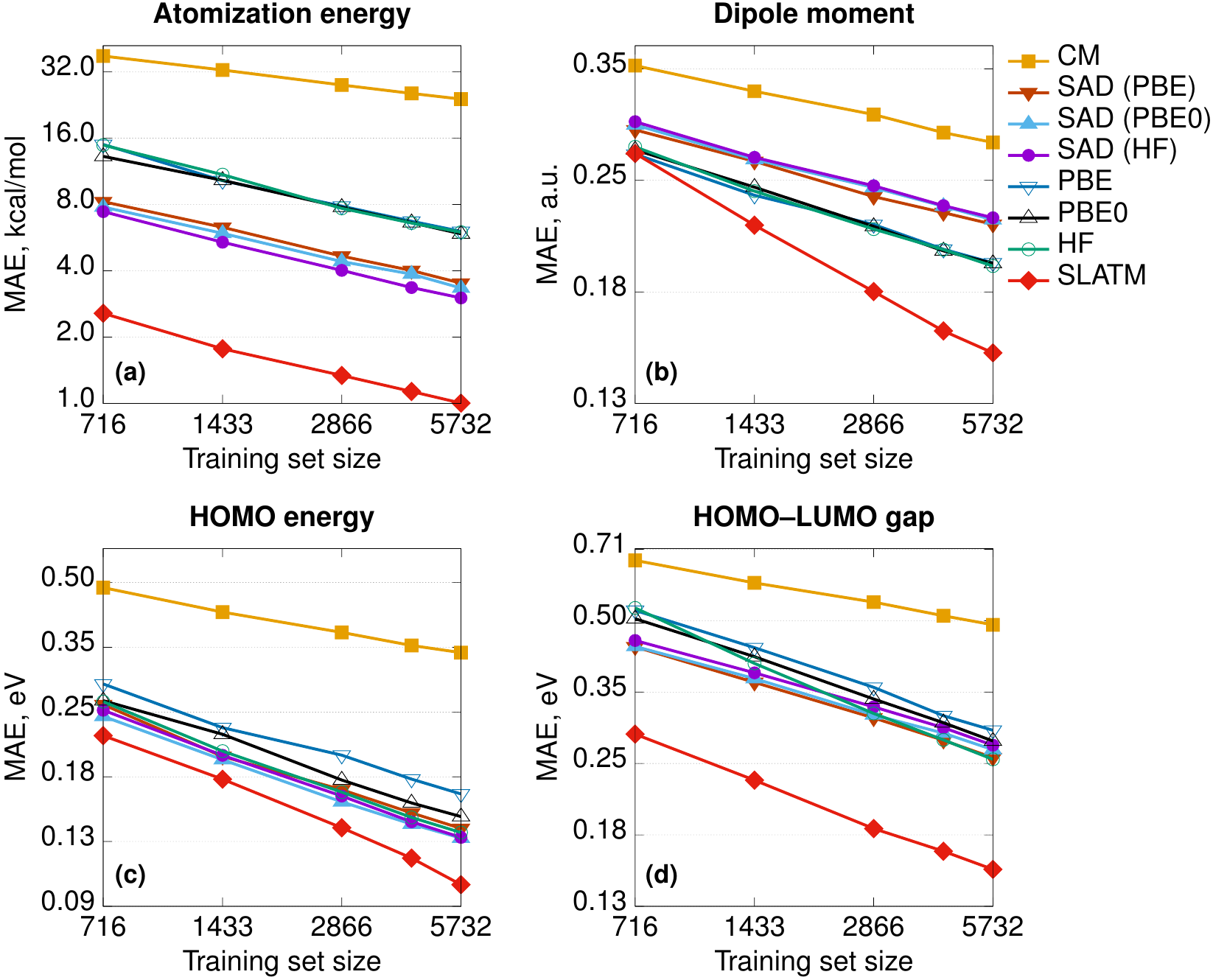}
\caption{
Learning curves for \SPAHM based on the SAD guess and converged Fock matrices
with different potentials. The learning curves for CM and SLATM are shown for
comparison.
}
\label{fig:potential}
\end{figure*}

All the quantum chemical guesses used in \SPAHM pass through the construction
of an approximate Hamiltonian, except for SAD that gives directly a density
matrix.\cite{AC2001,LZDG2006} \SPAHM-SAD is, therefore, the only representation
that requires the user to choose a potential. In Figure~\ref{fig:potential}, we
show that although different potentials have a larger impact on the quality of
the regression than the basis set, the changes are never sufficient to
qualitatively change the behavior of the representations.

Interestingly, the simplest of the potentials, Hartree--Fock, leads to most
accurate learning both as \SPAHM-SAD or as converged Fock matrix. This result,
particularly evident in the case of the atomization energies, corroborates the
conclusion drawn in the main text: adding more physics is not always met with
favorable response by the machine learning algorithm. The worsening of the
learning curves from HF to density functionals suggests that the introduction
of an (approximated) electron correlation potential has the same effect of
sparsification of the data in the representation space as observed for the SCF
procedure.

\clearpage

\section{Metallic complexes, large molecules, and conformational diversity}

Relying directly upon the information contained in the electronic Hamiltonian, the
SPA\textsuperscript{H}M philosophy naturally satisfies the injectivity criterion of quantum machine
learning representations. In this work, the practical realization of the SPA\textsuperscript{H}M
concept takes the form of an eigenvalue spectrum, whose advantage is being naturally invariant
under the basic symmetries of physics and well-defined for a molecular representation. Nonetheless,
larger molecules and transition metal complexes are usually characterized by a more dense
spectrum than the small molecules included in the QM7 and L11 databases.

\begin{figure*}[h]
\centering
\includegraphics[width=0.4 \textwidth]{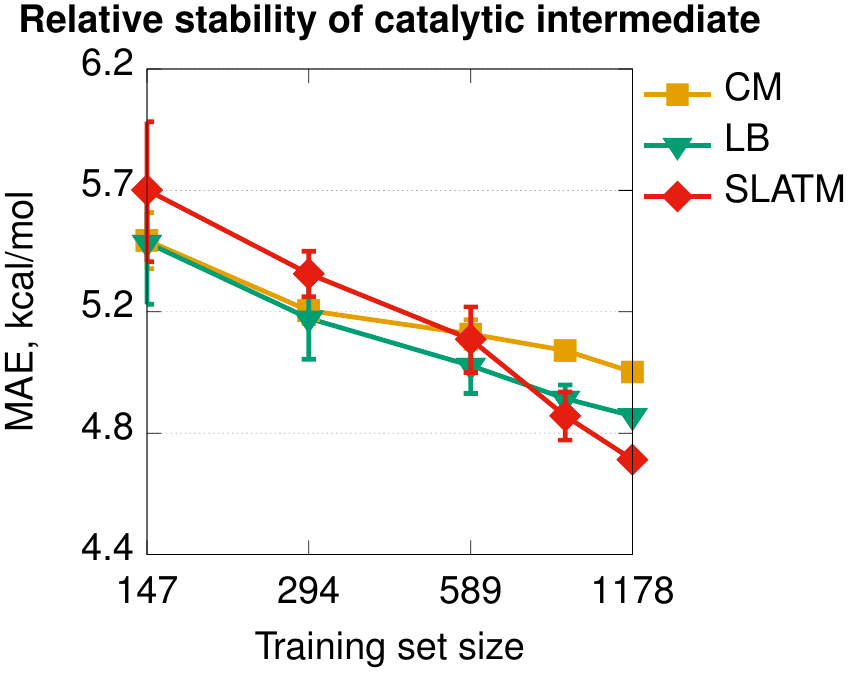}
\caption{
Learning curves for SPA\textsuperscript{H}M based on the LB Hamiltonian,
eigenvalue Coulomb Matrix (CM), and SLATM on a database of 1473~nickel complexes.
Regression target: energy descriptor [kcal/mol].
}
\label{fig:manuel}
\end{figure*}

\begin{figure*}[h]
\centering
\includegraphics[width=0.4 \textwidth]{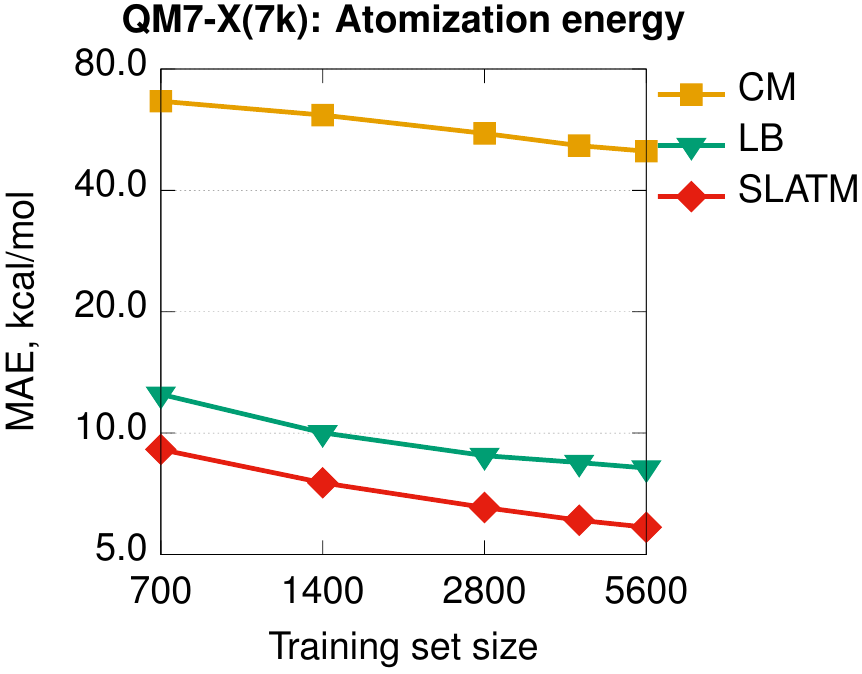}
\caption{
Learning curves for SPA\textsuperscript{H}M based on the LB Hamiltonian, eigenvalue Coulomb Matrix (CM),
and SLATM on the QM7-X database.\cite{HMEVDT2021}}
\label{fig:qm7x}
\end{figure*}

To test the reliability of eigenvalue SPA\textsuperscript{H}M representations on more complex
molecules, we report in Figure \ref{fig:manuel} the accuracy of the SPA\textsuperscript{H}M-LB
(LB), eigenvalue Coulomb Matrix (CM), and SLATM on a previously published database containing
1473~nickel complexes of varying size (from 22~atoms to 160~atoms).\cite{CWMSC2020} The organometallic database is freely available at  DOI:10.24435/materialscloud:fz-sw.
In addition, we test the applicability of eigenvalue SPA\textsuperscript{H}M
to learn the atomization energy among conformers (\ie constitutional and structural isomers and stereoisomers)
on the QM7-X database\cite{HMEVDT2021} (Figure~\ref{fig:qm7x}).

In both databases, the performance of
SPA\textsuperscript{H}M relative to existing representations is similar to the one reported in the
main text for QM7, particularly at full training set.

\clearpage

\section{Numerical Data}

\begin{table}[h]
\caption{Numerical values for the learning curves (Figure~\extref{1})}
\begin{ruledtabular}
\begin{tabular}{@{}ccccccccc@{}}
Training  &  \multicolumn{8}{c}{\bf Atomization energy MAE (SD), kcal/mol}        \\\cline{2-9}
set size  &   CM   &   $H_\mathrm{core}$ &   GWH  &     H\"uckel  &    SAP  &     SAD  &     LB  &      SLATM \\ \hline
 716 & 37.7    (0.7)    & 19.0    (0.5)   &  18.0    (0.4)    & 10      (1)      & 11.9    (0.7)   &  7.8       (0.4)  &  8.2     (0.4)   &  2.56    (0.03)    \\
1433 & 33      (1)      & 16.2    (0.4)   &  13.9    (0.3)    & 7.93    (0.08)   & 9.8     (0.3)   &  5.9       (0.3)  &  5.7     (0.3)   &  1.77    (0.02)    \\
2866 & 27.9    (0.4)    & 13.4    (0.1)   &  10.9    (0.2)    & 6.3     (0.1)    & 8.4     (0.3)   &  4.4       (0.1)  &  4.1     (0.2)   &  1.34    (0.03)    \\
4299 & 25.5    (0.2)    & 12.3    (0.1)   &  9.6     (0.1)    & 5.73    (0.04)   & 7.4     (0.1)   &  3.86      (0.07) &  3.7     (0.1)   &  1.13    (0.01)    \\
5732 & 24.0             & 11.5            &  8.8              & 5.25             & 6.9             &  3.34           &  3.2             &  1.00              \\
\hline \hline
\\
\hline \hline
Training   &  \multicolumn{8}{c}{\bf Dipole moment MAE (SD), kcal/mol}        \\\cline{2-9}
set size   &   CM   &   $H_\mathrm{core}$ &   GWH  &     H\"uckel  &    SAP  &     SAD  &     LB  &      SLATM \\ \hline
 716 & 0.357   (0.004)  & 0.343   (0.005) &  0.338   (0.003)  & 0.304   (0.008)  & 0.272   (0.005) &  0.297   (0.004) &  0.283   (0.002) &  0.272   (0.008)   \\
1433 & 0.330   (0.002)  & 0.320   (0.003) &  0.309   (0.005)  & 0.290   (0.005)  & 0.250   (0.007) &  0.267   (0.004) &  0.260   (0.005) &  0.218   (0.007)   \\
2866 & 0.307   (0.003)  & 0.303   (0.002) &  0.290   (0.003)  & 0.270   (0.003)  & 0.232   (0.003) &  0.245   (0.005) &  0.233   (0.002) &  0.177   (0.005)   \\
4299 & 0.290   (0.003)  & 0.295   (0.002) &  0.277   (0.001)  & 0.266   (0.003)  & 0.222   (0.002) &  0.231   (0.002) &  0.220   (0.004) &  0.157   (0.001)   \\
5732 & 0.281            & 0.286           &  0.270            & 0.258            & 0.217           &  0.221           &  0.211           &  0.146             \\
\hline \hline
\\
\hline \hline
Training   &  \multicolumn{8}{c}{\bf HOMO energy MAE (SD), eV}        \\\cline{2-9}
set size   &   CM   &   $H_\mathrm{core}$ &   GWH  &     H\"uckel  &    SAP  &     SAD  &     LB  &      SLATM \\ \hline
 716 & 0.49    (0.01)   & 0.47    (0.02)  &  0.41    (0.02)   & 0.36    (0.01)   & 0.295   (0.004) &  0.24    (0.02)  &  0.23    (0.01)  &  0.220   (0.009)   \\
1433 & 0.427   (0.009)  & 0.397   (0.008) &  0.35    (0.01)   & 0.300   (0.007)  & 0.273   (0.005) &  0.194   (0.006) &  0.186   (0.006) &  0.175   (0.003)   \\
2866 & 0.383   (0.009)  & 0.362   (0.005) &  0.310   (0.004)  & 0.257   (0.009)  & 0.238   (0.004) &  0.155   (0.002) &  0.157   (0.005) &  0.135   (0.002)   \\
4299 & 0.357   (0.003)  & 0.345   (0.003) &  0.288   (0.003)  & 0.230   (0.003)  & 0.223   (0.002) &  0.137   (0.001) &  0.140   (0.002) &  0.114   (0.003)   \\
5732 & 0.344            & 0.331           &  0.274            & 0.217            & 0.210           &  0.127           &  0.130           &  0.099             \\
\hline \hline
\\
\hline \hline
Training    &  \multicolumn{8}{c}{\bf HOMO--LUMO gap MAE (SD), eV}        \\\cline{2-9}
set size    &   CM   &   $H_\mathrm{core}$ &   GWH  &     H\"uckel  &    SAP  &     SAD  &     LB  &      SLATM \\ \hline
 716 & 0.67    (0.01)   & 0.63    (0.02)  &  0.56    (0.02)   & 0.54    (0.01)   & 0.478   (0.005) &  0.44    (0.01)  &  0.44    (0.01)  &  0.288   (0.013)   \\
1433 & 0.60    (0.01)   & 0.575   (0.007) &  0.483   (0.006)  & 0.458   (0.005)  & 0.43    (0.01)  &  0.38    (0.01)  &  0.38    (0.01)  &  0.231   (0.006)   \\
2866 & 0.55    (0.01)   & 0.520   (0.008) &  0.418   (0.006)  & 0.393   (0.006)  & 0.377   (0.004) &  0.316   (0.003) &  0.317   (0.005) &  0.182   (0.002)   \\
4299 & 0.51    (0.01)   & 0.491   (0.004) &  0.391   (0.003)  & 0.364   (0.005)  & 0.348   (0.005) &  0.290   (0.002  &  0.294   (0.004) &  0.163   (0.001)   \\
5732 & 0.49             & 0.474           &  0.368            & 0.344            & 0.324           &  0.267           &  0.277           &  0.150             \\
\end{tabular}
\end{ruledtabular}
\end{table}

\clearpage

\begin{table}[h]
\caption{Numerical values for the learning curves (Figure~\extref{2})}
\begin{ruledtabular}
\begin{tabular}{@{}cccccccc@{}}
\multirow{2}{*}{Training set size}
     &  \multicolumn{6}{c}{\bf Atomization energy MAE (SD), kcal/mol}        \\\cline{2-8}
     & LB
     & LB+0.001\,rnd
     & LB+0.005\,rnd
     & \multicolumn{1}{c}{LB+0.01\,rnd}
     & LB+0.1\,rnd
     & rnd
     & \multicolumn{1}{c}{PBE0}\\ \hline
716  & 8.2     (0.3) &    7.6     (0.5)   & 9.4     (0.9)  &   11      (2)     &  42      (1)      & 158     (2)    &   13.2    (0.6)\\
1433 & 5.6     (0.4) &    5.5     (0.2)   & 6.6     (0.3)  &   8.7     (0.2)   &  36.2    (0.6)    & 146.1   (0.8)  &   10.3    (0.6)\\
2866 & 4.2     (0.1) &    4.3     (0.1)   & 5.3     (0.2)  &   7.3     (0.2)   &  30.9    (0.3)    & 141.4   (1.0)  &   7.8     (0.4)\\
4299 & 3.6     (0.1) &    3.71    (0.07)  & 4.8     (0.1)  &   7.1     (0.2)   &  28.1    (0.4)    & 138.7   (0.6)  &   6.6     (0.1)\\
5732 & 3.2           &    3.30            & 4.5            &   6.6             &  25.9             & 138.3          &   5.9          \\
\end{tabular}
\end{ruledtabular}
\end{table}

\begin{table}[h]
\caption{Numerical values for the learning curves (Figure~\extref{3})}
\begin{minipage}{0.75\linewidth}
\begin{ruledtabular}
\begin{tabular}{@{}cccc@{}}
\multirow{2}{*}{Training set size}
       & \multicolumn{3}{c}{\bf Atomization energy MAE (SD), kcal/mol} \\\cline{2-4}
       &  core  & valence &  full   \\ \hline
716    &  94      (1)    &   18.9    (0.3)   &  8.2     (0.4)   \\
1433   &  86      (1)    &   15.2    (0.3)   &  5.7     (0.3)   \\
2866   &  81      (1)    &   12.3    (0.2)   &  4.1     (0.2)   \\
4299   &  77.3    (0.5)  &   10.6    (0.1)   &  3.7     (0.1)   \\
5732   &  75.3           &   9.7             &  3.2             \\
\hline \hline \\ \hline \hline
\multirow{2}{*}{Training set size}
       & \multicolumn{3}{c}{\bf Dipole moment MAE (SD), kcal/mol} \\\cline{2-4}
       &  core         & valence        &  full           \\ \hline
716    &  0.323 (0.009)&  0.296 (0.004) &  0.283  (0.002) \\
1433   &  0.297 (0.006)&  0.275 (0.006) &  0.260  (0.005) \\
2866   &  0.279 (0.002)&  0.255 (0.004) &  0.233  (0.002) \\
4299   &  0.272 (0.001)&  0.240 (0.004) &  0.220  (0.004) \\
5732   &  0.265        &  0.229         &  0.211          \\
\hline \hline \\ \hline \hline
\multirow{2}{*}{Training set size}
       & \multicolumn{3}{c}{\bf HOMO energy MAE (SD), eV} \\\cline{2-4}
       &  core  & valence &  full   \\ \hline
716    &  0.409    (0.007)& 0.218   (0.005)  & 0.23    (0.01)  \\
1433   &  0.363    (0.005)& 0.184   (0.006)  & 0.186   (0.006) \\
2866   &  0.339    (0.003)& 0.158   (0.003)  & 0.157   (0.005) \\
4299   &  0.325    (0.003)& 0.144   (0.001)  & 0.140   (0.002) \\
5732   &  0.314           & 0.134            & 0.130           \\
\hline \hline \\ \hline \hline
\multirow{2}{*}{Training set size}
       & \multicolumn{3}{c}{\bf HOMO--LUMO gap MAE, eV} \\\cline{2-4}
       &  core  & valence &  full   \\ \hline
716    &  0.60     (0.01) & 0.49    (0.02)   & 0.44    (0.01)  \\
1433   &  0.55     (0.01) & 0.43    (0.01)   & 0.384   (0.010) \\
2866   &  0.513    (0.008)& 0.363   (0.004)  & 0.317   (0.005) \\
4299   &  0.497    (0.003)& 0.329   (0.004)  & 0.294   (0.004) \\
5732   &  0.484           & 0.306            & 0.277           \\
\end{tabular}
\end{ruledtabular}
\end{minipage}
\end{table}

\begin{table}[h]
\caption{Numerical values for the learning curves (Figure~\extref{4}): HOMO energy MAE (SD), eV}
\begin{minipage}{0.75\linewidth}
\begin{ruledtabular}
\begin{tabular}{@{}ccccc@{}}
\multirow{2}{*}{Training set size}
       & \multicolumn{4}{c}{\bf $\bf M$ and $\bf M^{++}$ } \\\cline{2-5}
       & CM    & \SPAHM-LB & \SPAHM-LBm   & SLATM \\ \hline
720    &  6.38    (0.03) &    2.6     (0.1)  &   0.40    (0.01)  &  6.22   (0.03)  \\
1440   &  6.48    (0.02) &    1.7     (0.1)  &   0.337   (0.004) &  6.29   (0.02)  \\
2880   &  6.66    (0.01) &    1.13    (0.06) &   0.286   (0.003) &  6.30   (0.03)  \\
4320   &  6.73    (0.02) &    0.87    (0.03) &   0.260   (0.004) &  6.320  (0.006) \\
5760   &  6.85           &    0.70           &   0.244           &  6.334          \\
\hline \hline\\ \hline \hline
\multirow{2}{*}{Training set size}
       & \multicolumn{4}{c}{\bf $\bf M$ and $\bf M^{+\cdot}$ } \\\cline{2-5}
       & CM    & \SPAHM-LB & \SPAHM-LBm   & SLATM \\ \hline
720    &  3.35    (0.01)  &   1.3      (0.1)   &  0.35    (0.01)  &  3.27    (0.02)  \\
1440   &  3.38    (0.01)  &   0.86     (0.06)  &  0.31    (0.01)  &  3.27    (0.02)  \\
2880   &  3.46    (0.01)  &   0.60     (0.02)  &  0.273   (0.004) &  3.29    (0.01)  \\
4320   &  3.515   (0.008) &   0.498    (0.005) &  0.257   (0.004) &  3.298   (0.004) \\
5760   &  3.569           &   0.457            &  0.242           &  3.302           \\
\hline \hline\\ \hline \hline
\multirow{2}{*}{Training set size}
       & \multicolumn{4}{c}{\bf $\bf M$, $\bf M^{+\cdot}$, and $\bf M^{++}$ } \\\cline{2-5}
       & CM    & \multicolumn{1}{c}{\SPAHM-LB} & \SPAHM-LBm   & SLATM \\ \hline
1080   &  4.76    (0.02) &   1.87    (0.09) &   0.400   (0.008)  &   4.52    (0.01) \\
2160   &  4.71    (0.03) &   1.21    (0.04) &   0.340   (0.006)  &   4.47    (0.02) \\
4320   &  4.71    (0.01) &   0.76    (0.03) &   0.295   (0.003)  &   4.40    (0.02) \\
6480   &  4.70    (0.01) &   0.53    (0.02) &   0.276   (0.002)  &   4.38    (0.01) \\
8640   &  4.68           &   0.41           &   0.260            &   4.36           \\
\end{tabular}
\end{ruledtabular}
\end{minipage}
\end{table}

\begin{table}[h]
\caption{Numerical values for the learning curves (Figure~\extref{5}): dipole moment MAE (SD), a.u.}
\begin{minipage}{0.4\linewidth}
\begin{ruledtabular}
\begin{tabular}{@{}ccc@{}}
Training set size
     & SLATM        & \SPAHM-LB(m)  \\ \hline
 557 & 0.94  (0.06) & 0.71  (0.02)  \\
1111 & 0.79  (0.02) & 0.65  (0.01)  \\
2223 & 0.68  (0.01) & 0.588 (0.008) \\
4334 & 0.62  (0.01) & 0.561 (0.008) \\
5446 & 0.600        & 0.520         \\
\end{tabular}
\end{ruledtabular}
\end{minipage}
\end{table}

\begin{table}[h]
\caption{Numerical values for Figure~\extref{6}}
\begin{minipage}{0.85\linewidth}
\begin{ruledtabular}
\begin{tabular}{@{}lcccc@{}}
                                        & \multicolumn{2}{c}{\bf QM7}     & \multicolumn{2}{c}{\bf L11}     \\ \cline{2-3} \cline{4-5}
                                        & SLATM          & \SPAHM-LB(m)   & SLATM           & \SPAHM-LB(m)  \\ \hline
time to compute the representation, s   & \num{5.38E01}  & \num{4.79E02}  & \num{1.55E03}   & \num{8.01E01} \\
time to compute the kernel, s           & \num{5.38E01}  & \num{2.07E00}  & \num{2.93E02}   & \num{1.37E00} \\
\end{tabular}
\end{ruledtabular}
\end{minipage}
\end{table}

\clearpage
\section*{References}

\bibliography{refs.bib}